\documentclass[runningheads,bookmarksdepth=2]{llncs}

\usepackage{iflang}

\usepackage[T1]{fontenc}

\usepackage{mdframed}

\usepackage{graphicx}

\usepackage[svgnames,dvipsnames]{xcolor}

\usepackage{enumitem}

\usepackage{xspace}

\usepackage{tikz}
\usetikzlibrary{arrows.meta}

\usepackage{hyperref}
\hypersetup{
  bookmarksopen=true,    
  pdftitle={Making a Pipeline Production-Ready: Challenges and Lessons Learned in the Healthcare Domain},
  pdfauthor={D. Lawand et al.},
  pdfcreator={pdflatex}, 
  pdfnewwindow=true,     
  colorlinks=true,       
  linkcolor=blue,        
  citecolor=DarkGreen,   
  filecolor=green,       
  urlcolor=DarkRed       
}

\usepackage[nameinlink,noabbrev]{cleveref}

\usepackage[frozencache,cachedir=minted]{minted}

\newcommand{\etal}{\textit{et al.}\xspace}

\newcommand{\SPIRA}{\textsc{SPIRA}\xspace}

\newcommand{\Requirement}[1]{\textit{#1}\xspace}

\newcommand{\NFR}[1]{\Requirement{#1}}

\newcommand{\Pattern}[1]{\textsc{#1}\xspace}
\newcommand{\Smell}[1]{\textsc{#1}\xspace}

\newcommand{\SpiraV}[1]{\hyperref[sec:spira_v#1]{v#1}}

\definecolor{drawio-blue}{HTML}{1ba1e2} %
\definecolor{drawio-gray}{HTML}{647687} %
\definecolor{drawio-green}{HTML}{6d8764} %
\definecolor{drawio-orange}{HTML}{fa6800} %
\definecolor{drawio-pink}{HTML}{99004D} %
\definecolor{drawio-purple}{HTML}{76608a} %
\definecolor{drawio-red}{HTML}{e51400} %
\definecolor{drawio-white}{HTML}{f9f7ed} %
\definecolor{drawio-black}{HTML}{36393d} %
\definecolor{drawio-yellow}{HTML}{e3c800} %

\definecolor{drawio-violet}{HTML}{6a00ff} %
\definecolor{drawio-magenta}{HTML}{dd0073} %
\definecolor{drawio-moss}{HTML}{008a00} %

\newcommand{\TikzSkewedSquare}[1][black]{%
  \tikz[baseline] {\draw[transform shape, black, fill={#1}] (0,0) rectangle (1.5ex,1.5ex);}%
}
\newcommand{\TikzSkewedCircle}[1][black]{%
  \tikz[baseline] {\draw[yshift=0.75ex, anchor=south, transform shape, black, fill={#1}] (0,0) circle (0.75ex);}%
}

\newcommand{\LegendColoredSquare}[3]{%
  \textcolor{#1}{#3}~\TikzSkewedSquare[#2]%
}
\newcommand{\LegendColoredCircle}[3]{%
  \textcolor{#1}{#3}~\TikzSkewedCircle[#2]%
}

\newcommand{\LegendColoredComponent}[2]{%
  \LegendColoredSquare{drawio-#1}{drawio-#1}{#2}%
}
\newcommand{\LegendBWComponent}[1]{%
  \LegendColoredSquare{drawio-black}{drawio-white}{#1}%
}
\newcommand{\LegendColoredLabel}[2]{%
  \LegendColoredCircle{drawio-#1}{drawio-#1}{#2}%
}

\newcommand{\LegendService}[2]{%
  \textcolor{drawio-violet}{{#2~(#1)}}%
}
\newcommand{\LegendPipeline}[2]{%
  \textcolor{drawio-moss}{{#2~(#1)}}%
}
\newcommand{\LegendDataStore}[2]{%
  \textcolor{drawio-magenta}{{#2~(#1)}}%
}

\newmintedfile[InputAlgorithmLeft]{python}{
  numbers=left,
  numbersep=0.5em,
  fontsize=\tiny,
  frame=lines,
  framesep=0.5em,
  highlightcolor=Violet!10,
  labelposition=topline
}
\newmintedfile[InputAlgorithmRight]{python}{
  numbers=right,
  numbersep=0.5em,
  fontsize=\tiny,
  frame=lines,
  framesep=0.5em,
  highlightcolor=Violet!10,
  labelposition=topline
}

\emergencystretch=\maxdimen
\begin{document}



\title{%
Making a Pipeline Production-Ready:\break%
Challenges and Lessons Learned\break%
in the Healthcare Domain%
\vspace{-0.3cm}
}
\titlerunning{Making a Pipeline Production-Ready: Challenges and Lessons Learned}


\author{
Daniel Angelo Esteves Lawand\inst{1}\orcidID{0009-0007-0808-3241} \and \\
Lucas Quaresma Medina Lam\inst{1}\orcidID{0009-0002-0537-3095} \and \\
Roberto Oliveira Bolgheroni\inst{1}\orcidID{0009-0008-6663-9911} \and \\
Renato Cordeiro Ferreira\inst{1,2}\orcidID{0000-0001-7296-7091} \and \\
Alfredo Goldman\inst{1}\orcidID{0000-0001-5746-4154} \and \\
Marcelo Finger\inst{1}\orcidID{0000-0002-1391-1175}
}

\authorrunning{D. Lawand et al.}


\institute{%
Instituto de Matemática e Estatística (IME), University of São Paulo (USP),
Brazil \and
Jheronimus Academy of Data Science (JADS), Tilburg University~(TiU) and
Technical University of Eindhoven~(TUe), The Netherlands \\
\email{\{renatocf,gold,mfinger\}@ime.usp.br}
}


\maketitle 


\begin{abstract}
\vspace{-0.2cm}
Deploying a Machine Learning (ML) training pipeline into production
requires good software engineering practices.
Unfortunately, the typical data science workflow often leads to code
that lacks critical \mbox{software} quality attributes.
This experience report investigates this problem in \SPIRA,
a project whose goal is to create an ML-Enabled System (MLES)
to pre-diagnose insufficiency respiratory via speech analysis.
This paper presents an overview of the architecture of the MLES,
then compares three versions of its \nameref{subsec:spira_continuous_training}
subsystem:
  from a proof of concept \Pattern{Big Ball of Mud}~(\SpiraV{1}),
  to a design pattern-based \Pattern{Modular Monolith}~(\SpiraV{2}),
  to a test-driven set of \Pattern{Microservices}~(\SpiraV{3})
Each version improved its overall \NFR{extensibility}, \NFR{maintainability},
\NFR{robustness}, and \NFR{resiliency}.
The paper shares challenges and lessons learned in this process,
offering insights for researchers and practitioners seeking to
productionize their pipelines.

\keywords{%
Code Quality
\and MLOps
\and Software Architecture
\and Machine Learning Enabled Systems
\and Healthcare Domain
\and Experience Report
}%

\end{abstract}


\section{Introduction}
\label{sec:introduction}

In 2020, amidst the COVID-19 pandemic, a multidisciplinary team of
researchers from the University of São Paulo (USP) created \SPIRA
\footnote{\url{https://github.com/spirabr}}:
a project to detect respiratory insufficiency via speech analysis, using
Machine Learning (ML)~\cite{Finger2021DetectingProject}.

Since then, the scope of the \SPIRA project has evolved to detect respiratory
insufficiency of different origins, including many sicknesses that can cause
this symptom: smoking side effects, flu, severe asthma, and heart
conditions~\cite{Finger2021DetectingProject}.


To create a tool that could assist physicians, the team proposed to develop
the \SPIRA ML-Enabled System (MLES). Since 2020, different components have
been incrementally developed by bachelor students working with the project.


\section{The SPIRA ML-Enabled System}
\label{sec:the_spira_system}

\begin{figure}[p]
  \centering
  \includegraphics[width=0.98\linewidth]{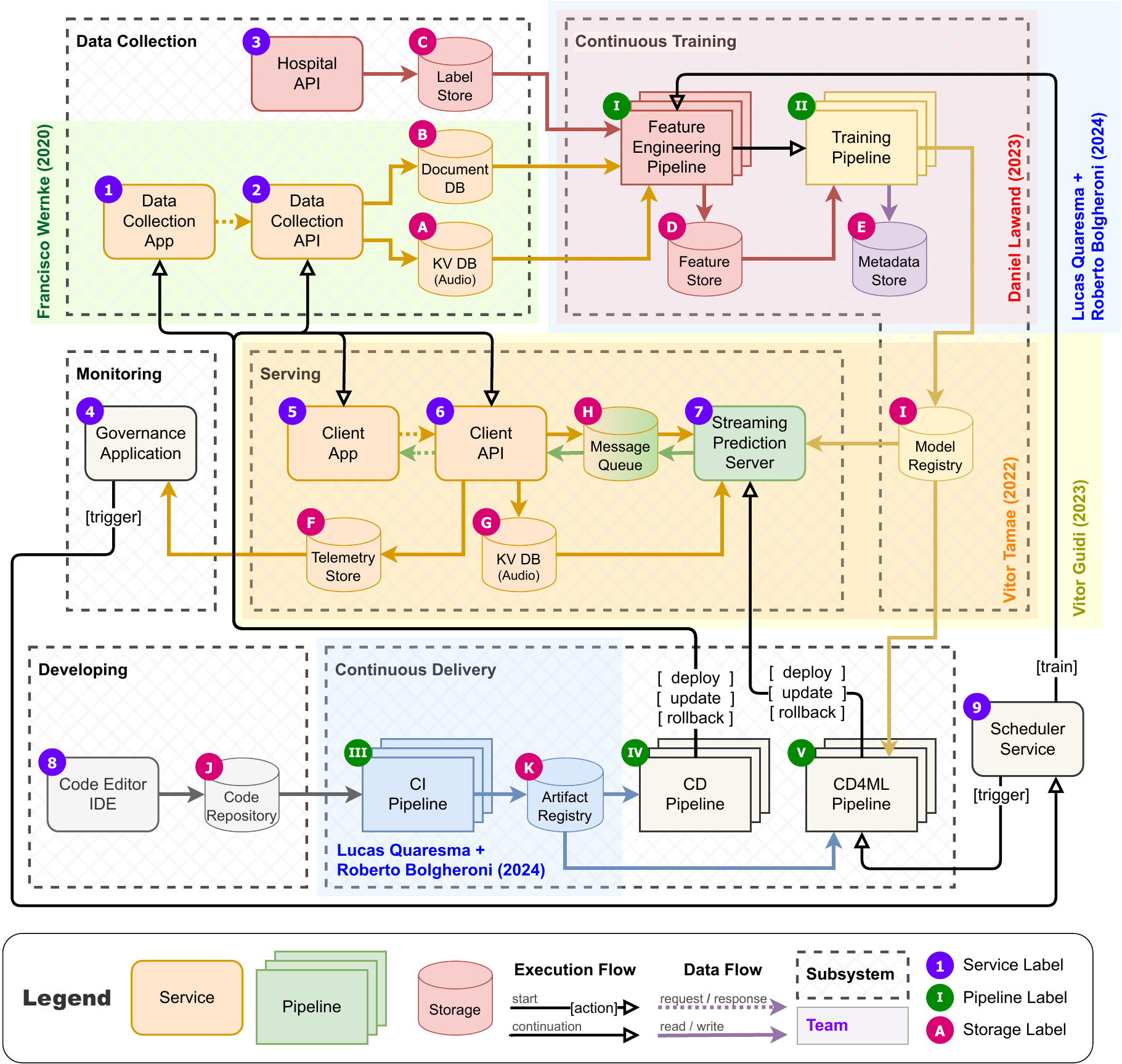}
  \caption[%
    Architecture of the SPIRA ML-Enabled System%
  ]{%
    \emph{Architecture of the SPIRA ML-Enabled System.}
    The architecture is described with the same notation of the
    reference architecture described in Ferreira \etal%
    ~\cite{Ferreira2025ASystems}.
    Rectangles represent \textbf{services},
      which execute continuously.
    Stacked rectangles represent \textbf{pipelines},
      which execute a task on demand.
    Lastly, cylinders represent \textbf{data storage},
      which may be databases of any type.
    Components are connected by arrows.
    Black arrows with a hollow tip illustrate the \textbf{execution flow}.
      They start and end in a component.
      Labeled arrows represent the trigger that starts a workflow,
      whereas unlabeled arrows represent the continuation of an
      existing workflow.
    Colored arrows with a filled tip illustrate the \textbf{data flow}.
    They appear in two types:
      solid arrows going to and from a data storage represent
      write and read operations, respectively;
      dotted arrows represent a sync or async request-response
      communication between components.
    Components are colored according to the data they produce:
      \mbox{\LegendColoredComponent{orange}{raw data}},
      \mbox{\LegendColoredComponent{gray}{source code}},
      \mbox{\LegendColoredComponent{blue}{executable artifacts}},
      \mbox{\LegendColoredComponent{red}{ML-specific data}},
      \mbox{\LegendColoredComponent{yellow}{ML models}},
      \mbox{\LegendColoredComponent{purple}{ML training metadata}},
      \mbox{\LegendColoredComponent{green}{ML model predictions}}, and
      \mbox{\LegendColoredComponent{pink}{ML model metrics}}.
    The remaining \LegendBWComponent{standalone components} orchestrate
    the execution of others.
    Components are also grouped into \textbf{subsystems},
    with their background colored according to the \textbf{students}
    responsible for their development.
    \LegendColoredLabel{violet}{Numbers},
    \LegendColoredLabel{moss}{roman numerals} and
    \LegendColoredLabel{magenta}{letters}
    are used as labels throughout \cref{sec:the_spira_system}.
  }
  \label{fig:spira_architecture}
\end{figure}

This section describes the architecture of the \SPIRA ML-Enabled System,
as illustrated by \cref{fig:spira_architecture}.
\Crefrange{subsec:spira_data_collection}{subsec:spira_monitoring}
showcase its six subsystems according to the reference architecture extended by
Ferreira \etal~\cite{Ferreira2025ASystems}.


  \vspace{-0.1cm} 
  \subsection{Data Collection}\label{subsec:spira_data_collection}

  Creating an ML model that can detect respiratory insufficiency
  via voice is a supervised machine learning classification problem%
  ~\cite{Casanova2021DeepSpeech}. As such, it requires \emph{labeled data}.
  The \SPIRA team supports a \LegendService{1}{Data Collection App} provided
  to volunteer data collectors to collect voices inside hospitals, where
  people with respiratory insufficiency may be found. The \LegendService{1}{
  Data Collection App} sends data to a \LegendPipeline{2}{Data Collection API},
  which in turn stores audio in an \LegendDataStore{A}{Audio Key-Value
  Database}, and saves patients' info in a \LegendDataStore{B}{%
  Document Database}.
  
  After each data collection, data collectors register a unique identifier for
  each participant. This ID can be cross-referenced with data shared by partner
  hospitals via 3rd-party \LegendService{3}{Hospital APIs}. This way,
  researchers can create a \LegendDataStore{C}{Label Store} that provides a
  \emph{ground truth} for training models. 

  \vspace{-0.1cm} 
  \subsection{\mbox{Continuous} \mbox{Training}}
  \label{subsec:spira_continuous_training}

  Casanova \etal~proposed a Deep Learning (DL) architecture based on a
  Convolutional Neural Network (CNN) to detect respiratory insufficiency
  from audio signals~\cite{Casanova2021DeepSpeech}. As more data becomes
  available as \emph{ground truth}, there is potential to retrain the
  model and improve its accuracy~\cite{Goodfellow:DeepLearning:2016}.
  

  The proof-of-concept shared by Casanova \etal~\cite{Casanova2021DeepSpeech}
  can be divided into two components:
    a \LegendPipeline{I}{Feature Engineering Pipeline},
    to process audio signals for the training algorithm; and
    a \LegendPipeline{II}{Training Pipeline},
    to train and validate the CNN.
  
  The development of these two pipelines is an opportunity to
  introduce standard machine learning design patterns%
  ~\cite{Lakshmanan:MLDesignPatterns:2020,Ferreira2025ASystems}, such as
    a \LegendDataStore{D}{Feature Store},
    a \LegendDataStore{E}{Metadata Store}, and
    a \LegendDataStore{I}{Model Registry}.
  For the \SPIRA project, the goal is to use \href{https://mlflow.org}{MLFlow}
  to make the two latter roles. On the other hand, for the features, the proposal
  is to use a simple key-value database like \href{https://min.io}{MinIO}.

  \vspace{-0.1cm} 
  \subsection{Development}\label{subsec:spira_development_ci}
  
  As more data becomes available as \emph{ground truth}, there is potential to
  redesign the architecture of the model used by \SPIRA to improve its accuracy%
  ~\cite{Goodfellow:DeepLearning:2016}. To make experimentation easy, it is
  desirable to provide a standardized environment for data scientists,
  compatible with production environments~\cite{Wilson:MLEInAction:2022}.

  Currently, all \SPIRA source code is open source in a
  \LegendDataStore{J}{Code Repository} at \href{https://github.com/spirabr}{GitHub}.
  Following the principles of \emph{Infrastructure as Code (IaC)}%
  ~\cite{Kief:InfrastructureAsCode:2025}, the \SPIRA organization
  provides standard configurations for popular development tools used
  by data scientists and machine learning engineers. In particular, this
  includes a basic setup for \LegendService{8}{Code Editors} (such as
  \href{https://code.visualstudio.com/}{VSCode}) or \LegendService{8}{
  IDEs} (such as \href{https://www.jetbrains.com/pycharm/}{PyCharm}),
  but can also include other utility scripts helpful for the developers.
  
  \vspace{-0.1cm} 
  \subsection{Continuous Delivery}
  \label{subsec:spira_continuous_delivery}

  To support the creation of new versions of the MLES -- after retraining
  or redesigning a model -- automation is essential to make deployment
  seamless.
  
  Taking advantage of GitHub, a \LegendService{III}{Continuous
  Integration Pipeline} uses
  \href{https://github.com/features/actions}{GitHub Actions} to generate
  \href{https://www.docker.com/}{Docker} containers for components,
  which are then deployed into the
  \href{https://github.blog/2020-09-01-introducing-github-container-registry/}{GitHub
  Container Registry}, the de facto \LegendDataStore{K}{Artifact Store} for
  \SPIRA.
  Conversely, the \LegendPipeline{IV}{Continuous Delivery Pipeline} and
  \LegendPipeline{V}{Continuous Delivery for Machine Learning Pipeline}
  use infrastructure configuration to deploy and rollback services in a
  \href{https://kubernetes.io/}{Kubernetes}-managed cluster.

  \vspace{-0.1cm} 
  \subsection{Serving}\label{subsec:spira_serving}

  Similarly to the \LegendService{1}{Data Collection App}, the
  \SPIRA \LegendService{5}{Client App} is meant to be used inside hospitals.
  However, it has a different goal: to provide (indirect) access to the
  \SPIRA model, pre-diagnosing respiratory insufficiency via speech analysis.
  Therefore, its data collection is more critical.

  Hospitals can have poor, unreliable internet reception. Multiple audios
  can be sent in a single burst whenever a user gets internet access.
  To prevent data loss, the \LegendService{5}{Client App} must store all data
  collected locally before sending it to its corresponding
  \LegendService{6}{Client API}. Moreover, to prevent data inconsistency,
  the \LegendService{6}{Client API} must validate all data received
  with the corresponding \LegendService{5}{Client App} before storing it
  in its \LegendDataStore{G}{Audio Key-Value Database}. 
  
  Deep Learning models such as the one proposed by Casanova \etal
  often require accelerator hardware (GPUs) to run efficiently%
  ~\cite{Goodfellow:DeepLearning:2016}. Since executing the model is a
  costly operation, handling arbitrarily bursts of request
  is difficult.
  
  To avoid this bottleneck, the \SPIRA ML model can be executed in a separate
  \LegendService{7}{Streaming Prediction Service}, decoupled from the
  \LegendService{6}{Client API}. A \LegendDataStore{H}{Message Queue} stores
  prediction requests and corresponding results. Thanks to this separation,
  the \LegendService{7}{Streaming Prediction Service} may be run in
  machines with GPUs, whereas other components are deployed in cheaper hardware.

  \vspace{-0.1cm} 
  \subsection{Monitoring}\label{subsec:spira_monitoring}

  Once the \SPIRA MLES reaches production, it needs to be maintained in
  operation. The \LegendDataStore{F}{Telemetry Store} stores useful logs and
  data about its usage, while the \LegendService{4}{Governance Application}
  summarizes statistics of its working status based on them. Combined, these
  two components help the \SPIRA team to decide to retrain the ML model,
  or consider redesigning the ML model, or update and redeploy a component
  by triggering the \LegendService{9}{Scheduler Service}.

\section{Incremental Implementation}
\label{sec:incremental_implementation}

\begin{figure}[p]
  \centering
  \includegraphics[width=0.99\linewidth]{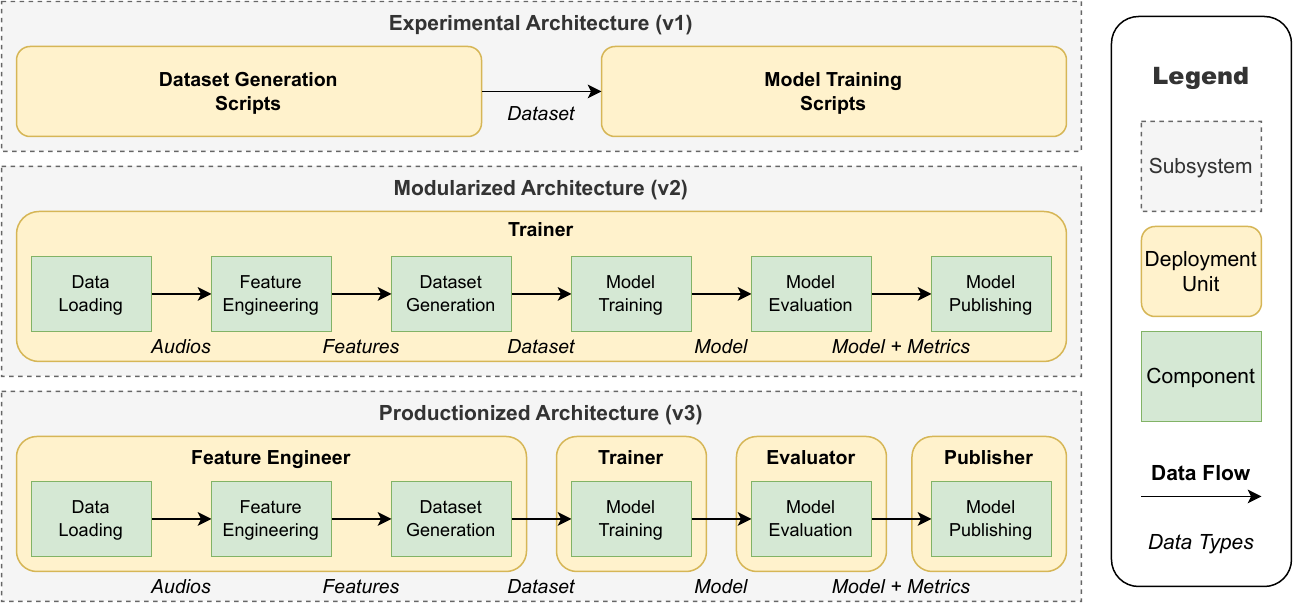}
  
  \vspace{-0.3cm} 
  \caption[%
    Incremental Development of the \SPIRA Continuous Training Subsystem.%
  ]{%
    \emph{Incremental Development of the SPIRA \nameref{subsec:spira_continuous_training} Subsystem.}
    \mbox{Dashed boxes~\TikzSkewedSquare[gray]}
    represent the versions of the subsystem.
    \mbox{Rounded boxes \TikzSkewedSquare[yellow]}
    represent deployment units.
    \mbox{Squared boxes \TikzSkewedSquare[green]}
    represent components.
    Black arrows with a filled tip illustrate the \textbf{data flow},
    while the labels in \textit{italic} represent the \textbf{data types}.
  }
  \label{fig:spira_evolution}
\end{figure}

The \SPIRA MLES has been incrementally developed since 2021.
This paper focuses on the \nameref{subsec:spira_continuous_training} subsystem,
whose development process is summarized in \cref{fig:spira_evolution}.
For challenges and lessons learned while developing the
\nameref{subsec:spira_data_collection} and \nameref{subsec:spira_serving},
subsystems, please refer to our paper \emph{``SPIRA: Building an Intelligent
System for Respiratory Insufficiency Detection''}~\cite{Ferreira2022SPIRA:Detection}.

\section{Experimental Architecture (v1)}
\label{sec:spira_v1}

The first version of the \nameref{subsec:spira_continuous_training} pipeline
was created as a proof-of-concept to showcase the \SPIRA model. The code was
published as part of the \href{https://doi.org/10.5281/zenodo.15824043}{%
reproduction package} by Casanova \etal~in 2020~\cite{Casanova2021DeepSpeech}.

The package has two main sets of scripts: one to create datasets (available
in the \texttt{scripts} folder), and one to train the CNN model (available
in the \texttt{models} folder). The code presents an unplanned, organically
developed structure, as it often occurs in the experimentation phase by
data scientists~\cite{Lakshmanan:MLDesignPatterns:2020,Wilson:MLEInAction:2022}.

\hyperref[fig:improving_modularization]{Figure 3 (left)} illustrates an example
of the \Smell{Misplaced Responsibility} bad smell~\cite{Martin:CleanCode:2008}
in the code, evidence of high coupling and low cohesion in the code.
In this way, this architecture can be classified as a \Pattern{Big Ball of
Mud}~\cite{Foote1999BigMud}.

\begin{figure}[p]
\center
\begin{minipage}{0.496\linewidth}
  \InputAlgorithmLeft[
    label=( v1 architecture ),
    highlightlines={ 1-6, 8-16, 18-19, 22-24, 30 },
  ]{code/misplaced_responsability.py}
  \label{subfig:misplaced_responsability}
\end{minipage}
\begin{minipage}{0.496\linewidth}
  \InputAlgorithmRight[
    label=( v2 architecture ),
    highlightlines={ 3-7, 15-23, 27, 29-36 },
  ]{code/pipeline.py}
  \label{subfig:adding_patters}
\end{minipage}

\vspace{-0.5cm} 
\caption[
   \emph{Example of improving modularization between \SpiraV{1}
   and \SpiraV{2} architectures.}
]{
   \emph{Example of improving modularization between \SpiraV{1}
   and \SpiraV{2} architectures.}
 
   \vspace{0.1cm}
 
   The \SpiraV{1}~snippet (left) shows a
   \Smell{Misplaced Responsibility} bad smell~\cite{Martin:CleanCode:2008}
   at the class \texttt{Dataset}.
   Lines 1-7 handle random number generation.
   Lines 8-16 assign values to attributes.
   Lines 18-19 and 22-24 handle assertions.
   Line 30 handles data loading.

   \vspace{0.1cm}
 
   The \SpiraV{2}~snippet (right)
   shows the application of design patterns
   at the module \texttt{pipeline}.
   Line 15-23 handles data loading using the \texttt{Audio}
   \Pattern{Adapter}.
   Line 27 builds an \texttt{audio\_processor} via
   a \Pattern{Chain of Responsibility} pattern.
   Line 29-36 build \texttt{feature\_transformer}s via
   the \Pattern{Strategy} pattern.

   \vspace{0.1cm}
   
   Improving the modularization makes the intention of the code more
   explicit: allow multiple experiments depending on the configuration.
}
\label{fig:improving_modularization}
\vspace{-0.2cm}
\end{figure} 
\section{Modularized Architecture (v2)}
\label{sec:spira_v2}

The second version of the \nameref{subsec:spira_continuous_training}
pipeline was created during the \href{https://web.archive.org/web/*/https://danlawand.github.io/MAC0499-Capstone-Project/*}{bachelor thesis}
of Daniel Lawand~\cite{Lawand2023EnablingPipeline}.
Its code is open source at
\href{https://doi.org/10.5281/zenodo.15824053}{GitHub}.

This version applied multiple design patterns~\cite{Gamma:DesignPatterns:1994}
to modularize the code. Some examples include:
\Pattern{Chain of Responsibility},
  to dynamically choose the pipeline steps to be executed;
\Pattern{Strategy},
  to dynamically choose techniques for data preprocessing, feature engineering,
  and model evaluation; and
\Pattern{Template Method},
  to reuse generic code snippets.
In another level of abstraction, the business logic was decoupled from external
dependencies by using the \Pattern{Ports and Adapters} architectural pattern%
~\cite{Martin:CleanArchitecture:2017}, using \Pattern{Dependency Injection} to
connect different layers of the application.

\hyperref[fig:improving_modularization]{Figure 3 (right)} illustrates the use
of design patterns. This reimplementation made the code more \emph{maintainable}
and \emph{extensible}~\cite{Richards:FundamentalsSoftwareArchitecture:2020}.
The \SpiraV{2} architecture can be classified as a \Pattern{Modular
Monolith}~\cite{Richardson:MicroservicesPatterns:2018}.

\section{Productionized Architecture (v3)}
\label{sec:spira_v3}

The third version of the \nameref{subsec:spira_continuous_training}
pipeline was created during the
\href{https://web.archive.org/web/*/https://lucasqml.github.io/mac0499/*}{bachelor
thesis}
of Lucas Quaresma and Roberto Bolgheroni~\cite{Lam2024ProductionizingSystem}.
Its code is open source at
\href{https://doi.org/10.5281/zenodo.15824056}{GitHub}.

This version expands on the \SpiraV{2} architecture by splitting
the single monolithic application into multiple deployment units,
as described by the \Pattern{Workflow Pipeline} ML design pattern%
~\cite{Lakshmanan:MLDesignPatterns:2020}.
The goal was to enable executing the pipeline with a
\LegendService{9}{Scheduler Service}.
In this way, if a failure occurs in one stage of the pipeline, it does not
require re-executing it from the start, a costly operation.

Testability was also a key priority in this version. It was only possible
because the \SpiraV{2} architecture made clearer the expected behaviors
from the \SpiraV{1} architecture. Following testing practices for
microservices~\cite{Richardson:MicroservicesPatterns:2018},
the business logic received unit tests via \emph{Test-Driven Development}
(TDD)~\cite{Beck:TDDByExample:2002}, while the \Pattern{Ports and Adapters}
received integration tests around external
dependencies~\cite{Martin:CleanArchitecture:2017}.

This reimplementation made the code more \emph{robust} and \emph{resilient}%
~\cite{Richards:FundamentalsSoftwareArchitecture:2020}.
The \SpiraV{3} architecture follows a \Pattern{Microservices} architectural
style~\cite{Newman:BuildingMicroservices:2021}.
\section{Challenges}
\label{sec:challenges}
\vspace{-0.2cm} 

The \nameref{sec:spira_v1} was built following a typical \mbox{CRISP-DM}
process~\cite{Burkov:MachineLearningEngineering:2020,Wilson:MLEInAction:2022}.
The goal was to create a proof-of-concept model. Serving it, i.e.,
going into production, was the last step of the workflow.

This development process brought a series of drawbacks for the
\nameref{subsec:spira_continuous_training} subsystem's architectural
characteristics, in particular its
  \NFR{extensibility},
  \NFR{maintainability},
  \NFR{robustness}, and
  \NFR{resiliency}.
As a consequence, there were two main challenges to migrating the subsystem
toward production.

\vspace{-0.2cm} 
\paragraph{Separation of Concerns.}
\label{p:separation_of_concerns}
The \SpiraV{1} architecture had many examples of \Smell{Misplaced Responsibility}
bad smell~\cite{Martin:CleanCode:2008},
which affected its overall \NFR{maintainability} and \NFR{extensibility}%
~\cite{Richards:FundamentalsSoftwareArchitecture:2020}.
Building the \SpiraV{2} architecture became an exercise in
\emph{Software Archeology}~\cite{Hunt2002SoftwareArchaeology}:
it was reimplemented from the \SpiraV{1} code by carefully reading it line by line
to decipher its intentions, applying design patterns to decrease coupling and
increase cohesion in its abstractions.

\vspace{-0.2cm} 
\paragraph{Automated Testing.}
\label{p:automated_testing}
The \SpiraV{2} architecture had improved \NFR{modularity},
but it still lacked \NFR{robustness} and \NFR{resiliency}%
~\cite{Richards:FundamentalsSoftwareArchitecture:2020}.
Building the \SpiraV{3} architecture became an \mbox{exercise} in
\emph{Test-Driven Development} (TDD)~\cite{Beck:TDDByExample:2002}:
it was reimplemented from \SpiraV{2} code by carefully interpreting intentions
around its design, creating automated tests to document behaviors that should be
maintained for the long term.

\section{Lessons Learned}
\label{sec:lessons_learned}
\vspace{-0.2cm} 

The development of MLES is inherently difficult, and developing them without
adequate planning makes their complexity even greater. As a consequence,
there were two lessons learned that could have improved the migration of the
\nameref{subsec:spira_continuous_training} subsystems toward production.

\vspace{-0.2cm} 
\paragraph{Collaboration between Data Scientists and ML Engineers.}
\label{p:collab_ds_mle}
Machine Learning Engineering (MLE) is a new subarea of software engineering
whose goal if to help to productionize ML models%
~\cite{Burkov:MachineLearningEngineering:2020,Wilson:MLEInAction:2022}.
This discipline helps the data scientist to think about how it
will fit into an MLES.
By bringing these two roles together since the beginning,
it is easier to make code \NFR{maintainable} and \NFR{extensible}.

\vspace{-0.2cm} 
\paragraph{Testing as a First-Class Concern.}
\label{p:testing_first-class_concern}
Creating automated tests is recognized as a good practice in
software engineering~\cite{Beck:XP:2004}.
This discipline helps developers to think about the long-term
maintenance of a system.
By designing tests and validation since the beginning,
it is easier to make the code \NFR{robust} and \NFR{resilient}.

\section{Conclusion}
\label{sec:conclusion}
\vspace{-0.2cm} 

This paper addressed the incremental development of \SPIRA, focusing on its
\nameref{subsec:spira_continuous_training} subsystem.
Its architecture evolved in three stages:
  from a proof of concept \Pattern{Big Ball of Mud}~(\SpiraV{1}),
  to a design pattern-based \Pattern{\mbox{Modular} \mbox{Monolith}}~(\SpiraV{2}),
  to a test-driven set of \Pattern{Microservices}~(\SpiraV{3}).
Each step helped the subsystem's 
  \NFR{extensibility},
  \NFR{maintainability},
  \NFR{robustness}, and
  \NFR{resiliency}.

By learning from this experience, similar projects may employ the above lessons
learned to avoid similar challenges, thus reaching production sooner, 


\subsubsection*{Acknowledgements.} This research was funded by FAPESP
(São Paulo Research Foundation), grant number 2023/00488-5.


%
%
%
\bibliographystyle{splncs04}
\bibliography{mendeley}

\end{document}